# High-order harmonic generation from laser induced plasma comprising CdSe/V$_2$O$_5$ Core/Shell quantum dots embedded on MoS$_2$ nanosheets

Srinivasa Rao Konda,[1,*] Puspendu Barik,[1] Subshash Singh,[2] Venkatesh Mottamchetty,[1,3] Amit Srivasthava,[4] Vyacheslav V. Kim,[5] Rashid A. Ganeev,[6,7] Chunlei Guo,[2] and Wei Li [1,*]

[1] The GPL Photonics Laboratory, State Key Laboratory of Luminescence and Applications, Changchun Institute of Optics, Fine Mechanics and Physics, Chinese Academy of Sciences, Changchun, Jilin 130033, China
[2] The Institute of Optics, University of Rochester, Rochester, NY, 14627, USA
[3] Department of Materials Science and Engineering, Uppsala University, Box 35, SE-75103 Uppsala, Sweden
[4] Department of Physics, TDPG College, VBS Purvanchal University, Jaunpur, 222001, India
[5] Laboratory of Nonlinear Optics, University of Latvia, Jelgavas 3, Riga, LV – 1586, Latvia
[6] Institute of Theoretical Physics, National University of Uzbekistan, Tashkent 100174, Uzbekistan
[7] Department of Optics and Spectroscopy, Voronezh State University, 1 University Square, Voronezh 394006, Russia

Correspondence: *ksrao@ciomp.ac.cn, *weili1@ciomp.ac.cn

**Abstract**

Research of the nonlinear optical characteristics of transition metal dichalcogenides in the presence of photoactive particles, plasmonic nanocavities, waveguides, and metamaterials is still in its early stages. This investigation delves into the high-order harmonic generation (HHG) from laser induced plasma of MoS$_2$ nanosheets in the presence of semiconductor photoactive medium such as CdSe and CdSe/V$_2$O$_5$ core/shell quantum dots. Our comprehensive findings shed light on the counteractive coupling impact of both bare and passivated quantum dots on MoS$_2$ nanosheets, as evidenced by the emission of higher-order harmonics. Significantly, the intensity of harmonics and their cut-off were notably enhanced in the MoS$_2$-CdSe and MoS$_2$-V-CdSe configurations compared to pristine MoS$_2$ nanosheets. These advancements hold promise for applications requiring the emission of coherent short-wavelength radiation.

## Introduction

Nonlinear optics emphasizes the interaction of intense light with matter and is also one of the fundamental building blocks of modern optics. High-order harmonic generation (HHG) in laser-induced plasmas (LIP) is a strong-field nonlinear optical process aimed in formation of the coherent extreme ultraviolet (XUV) radiation sources, generation of attosecond pulses, and analysis of the influence of micro- and nanoparticles on the harmonics generation efficiency [1–3]. The transition metal dichalcogenides (TMDs), especially 2D TMDs, have demonstrated fascinating optical properties compared with graphene because of a sandwich-like structure of a transition metal atom positioned between two layers of chalcogen atoms, the optical band gap at room temperature, and thermodynamically stable structural phases, i.e., either trigonal prismatic (2H) or octahedral (1T) coordination of the metal atoms [4–7]. Moreover, 2D TMDs are attractive for HHG owing to their strong many-body interaction and distinctive electronic properties, making them essential in strong-field and attosecond physics.[8,9] Research in the new material synthesis-based approach for layered van-der-Waals materials to control optical nonlinearities and optimization in emission of higher-order harmonics either from solids or LIPs open the way to explore the stable XUV light sources.

In this article, we demonstrate CdSe (core) and CdSe/$V_2O_5$ (core/shell, termed as V-CdSe now on) quantum dots (QDs) embedded on the few-layer $MoS_2$ nanosheet to study HHG comprehensively in exquisite detail. We investigate the effect of pure CdSe QD and passivated core/shell QD (V-CdSe) for the efficient generation of high-order harmonics in LIP. In this respect, the experimental observation and understanding of counteractive effects of CdSe and V-CdSe during HHG using $MoS_2$ nanosheets is of fundamental interest and can lead to the knowledge of the role of QDs responsible for the enhanced HHG performance of 2D nanosheets.

## Materials and Methods

**Materials Synthesis:** The $MoS_2$ nanosheets were synthesized through a hydrothermal approach in a Teflon-based autoclave at 210 °C, adapted from a method reported earlier.[10] CdSe (~3 nm, core) and CdSe/$V_2O_5$ (~4.5 nm, core/shell) QDs were synthesized using the hydrothermal method, described in detail in previously reported work[11] The nanosheet-QDs composite structures were synthesized via the hydrothermal method by adding QDs, as briefly described in the references [10,11]

## Experiential details for high-order harmonic generation

For HHG experiments, the powdered samples were made in smaller palettes and attached to the glass slide. The samples were kept in the target chamber containing a 3D stage to adjust their position with regard to the focused femtosecond pulses. High-order harmonics were generated from the LIPs of $MoS_2$ nanosheets, $MoS_2$-CdSe, and $MoS_2$-V-CdSe ablated by either nanosecond (ns) or picosecond (ps) heating pulses (HP). HHG was analyzed during, single and two-color femtosecond (fs) driving pulses (DP) with different delays between ns HP and fs DP, and various intensities of fs DP and ps HP. A Ti: Sapphire laser operating at 800 nm (35 fs) with a repetition rate of 100 Hz was used during these studies for single color pump. For the two-color pump (800 nm+400 nm) we inserted BBO (type 1, 0.2 mm thick) on the path of the focused DP inside the target chamber as shown in Figure 1 [12]. It was measured that the conversion efficiency of fundamental beam was 8 % for second harmonic.

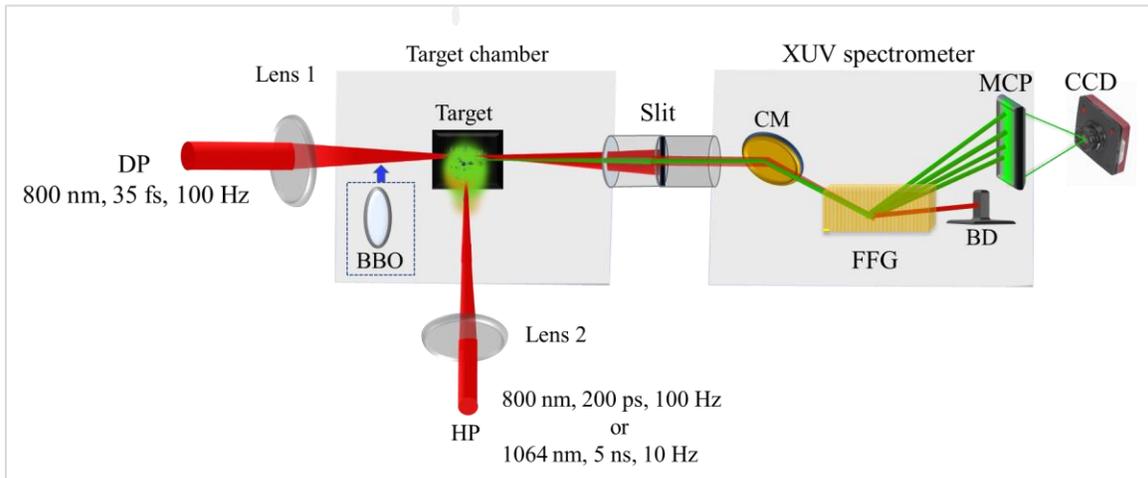

**Figure 1.** (a) Experimental layout for HHG. Driving pulse (DP) was focused by a spherical lens (Lens 1, f=500 mm) inside the plasma and lens 2 (f=200 mm) was used to create the plasma on targets. In the case of TCP, beta barium borate (BBO) was inserted on the path of focused radiation and produced laser plasma on the surface of the target. The XUV spectrometer comprising the cylindrical mirror (CM), flat-field grating (FFG), and microchannel plate (MCP). The residual driving beams was blocked by beam dumper (BD). A CCD camera registered the harmonic spectrum.

The harmonics spectra were measured using a single-color pump (800 nm) and two-color pump (800 nm+400 nm) with a 35 fs beam focused by a 500 mm focal length spherical lens (focal spot size of 43.5 µm). We used 200 ps (800 nm) and 5 ns (1064 nm, 10 Hz, Q-smart 850) pulses as HP to produce the plasma above the target samples by a 200 mm focal length spherical lens. The delay between ps HP and fs DP was fixed at 80 ns. In contrast, for ns HP, the time delay between HP and DP was controlled using a delay generator (Stanford Research Instrument, DG535). The emission of produced harmonics was detected by an XUV spectrometer comprised a cylindrical mirror, a flat field grating with 1200 grooves/mm, and a microchannel plate (MCP) with phosphor screen, which was used to magnify the XUV signal. A CCD camera captured images of harmonics appearing on the screen.

## Results and discussion

### HHG from plasma using single-color pump (800 nm):

*Plasma produced by nanoseconds pulses:*

Figure 2a-c shows the HHG spectra in the case of the single-color pump (with 800 nm), and Figure 2d shows the maximum harmonic signal achieved for $MoS_2$ nanosheets, $MoS_2$-CdSe, and $MoS_2$-V-CdSe in the range of 100 – 800 ns delays between HP and DP. The HHG spectra were recorded using the ns HP with an energy of 1.1 mJ and fs DP with an energy of 0.45 mJ (*I*: $3.2 \times 10^{14}$ W/cm$^2$). The inset in Figure 2d shows the harmonic cut-off for three samples. In the case of pristine $MoS_2$ nanosheets, the maximum harmonic signal was achieved at 100 ns delay, which was shifted to 200 ns for $MoS_2$-CdSe and $MoS_2$-V-CdSe. The maximum cut-offs observed for pure $MoS_2$ nanosheets, $MoS_2$-CdSe, and $MoS_2$-V-CdSe were 17H, 21H, and 25H, respectively. These cut-offs were consistent concerning the delay range between 100-600 ns, 100-700 ns, and 100-800 ns for corresponding individual samples (Figure 2a-c). The first three maxima appear in the case of pristine $MoS_2$ nanosheets for delays of 100 ns, 300 ns, and 500 ns, respectively (Figure 2d). The maximum peak position was achieved at a further 200 ns delay for samples.

The different maximum peak positions of harmonic intensity (Fig. 1d) indicate that plasma contains the fast and slow components comprising the individual atoms or ions and small to medium-sized clusters, respectively. We pointed out in our earlier work that the largest harmonic yield at each delay suggests that a significant number of the plasma particles arrive at the DP's focal point at that moment.[13] By dividing the lateral distance (0.2 µm) between the DP and the sample surface by the delay value, one may determine the velocity of this set of particles, which equals to $2 \times 10^3 \; m/s$ at 100 ns and $1 \times 10^3 \; m/s$ at 200 ns delay. For a given distance of 0.2 µm, the predicted velocities at different delay times are occupied by $2 \times 10^3 \; m/s$ to $0.25 \times 10^3 \; m/s$ between 100 and 800 ns delay. As the delay increased, the velocity of particles decreased, corresponding to a fixed position, which confirms the contribution of HHG from the clusters of target samples.

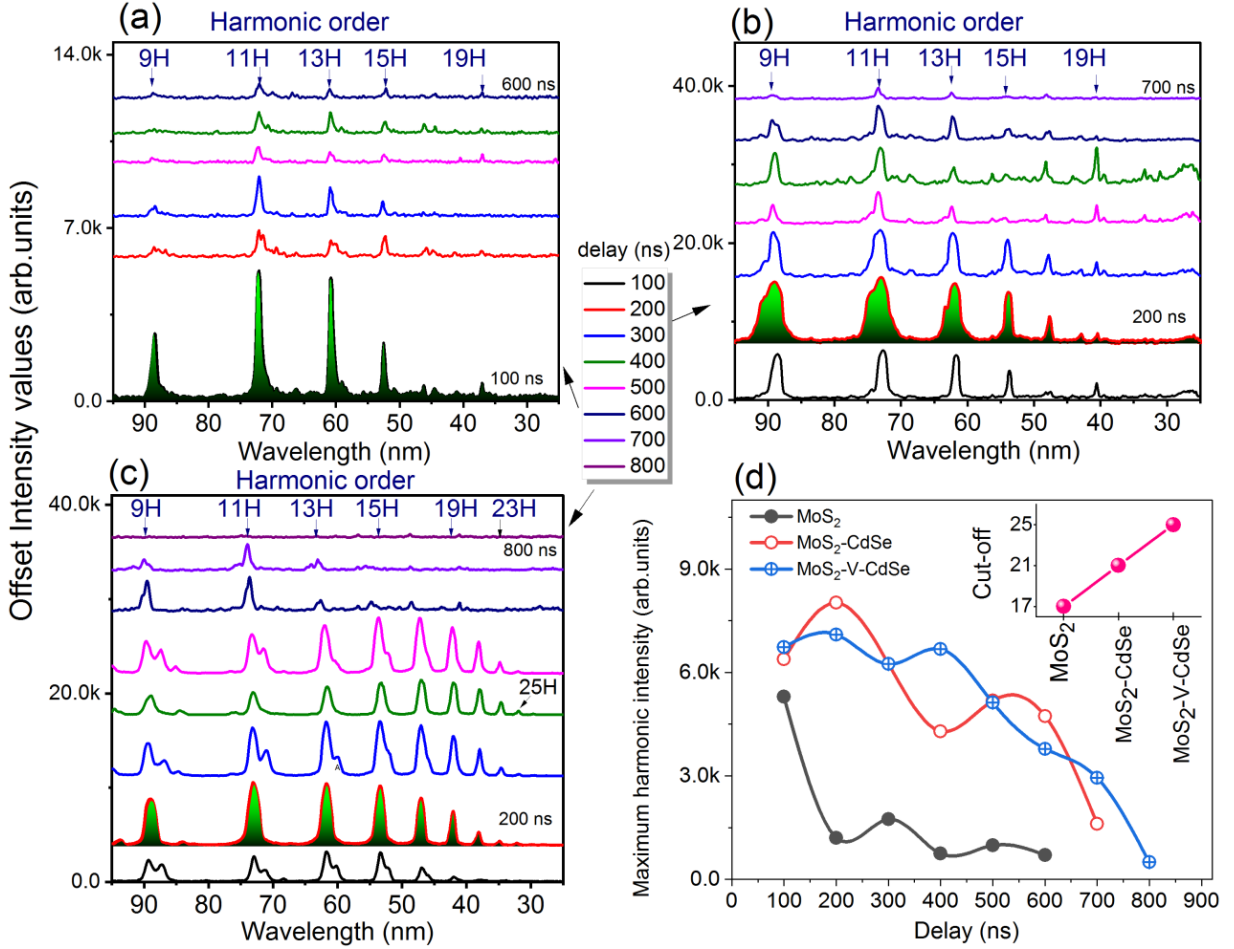

**Figure 2.** HHG spectra concerning delay between ns HP and fs DP (a) MoS$_2$ nanosheets, (b) MoS$_2$-CdSe, (c) MoS$_2$-V-CdSe. (d) the maximum harmonic intensity and inset shows the harmonic cut-off for three samples.

*Plasma produced by picosecond pulses*

Figure 3a shows the raw HHG spectra of MoS$_2$ nanosheets (top panel), MoS$_2$-CdSe (middle panel), and MoS$_2$-V-CdSe (bottom panel) for DP at 800 nm (35 fs). The measurements were carried out by varying fs DP energy and ps HP energy. The energy sets considered here (fs DP, ps HP) were (130 µJ, 36 µJ), (19 µJ, 46 µJ), (30 µJ, 60 µJ), (38 µJ, 85 µJ), (45 µJ, 120 µJ), and the corresponding intensities were (1.2×10$^{14}$ W/cm$^2$, 2.1×10$^9$ W/cm$^2$), (1.7×10$^{14}$ W/cm$^2$, 2.7×10$^9$ W/cm$^2$), (2.7×10$^{14}$ W/cm$^2$, 3.6×10$^9$ W/cm$^2$), (3.4×10$^{14}$ W/cm$^2$, 5.1×10$^9$ W/cm$^2$), and (4.1×10$^{14}$ W/cm$^2$, 7.1×10$^9$ W/cm$^2$), respectively.

Figure 3b compares harmonic spectra in terms of their intensity distribution at $E_{fs}$= 300 µJ and $E_{ps}$=60 µJ, which indicates the MoS$_2$-V-CdSe has a stronger harmonic intensity than other two samples. As shown in Figure 3a, the pristine MoS$_2$ nanosheets show that the harmonic emission started at 300 µJ. However, it starts at 190 µJ for MoS$_2$-CdSe and at 130 µJ for MoS$_2$-V-CdSe. The results indicate that the presence of CdSe and V-CdSe QDs enhances the plasma density and the components such as atoms and ions appear at higher density, which improves the harmonics intensity and harmonic cut-off as shown in Figures 3c and 3d. The harmonic cut-off can be higher at smaller DP intensity depending on the plasma consistence. Particularly, MoS$_2$-CdSe has a cut-off at 21H at 190 µJ while it is extended up to 29H for MoS$_2$-V-CdSe. However, the same cut-off of 29H is possible at DP energy 300 µJ for pristine MoS$_2$ nanosheets. The three samples have different numbers of elements - pristine MoS$_2$ nanosheet has two elements (i.e., Mo and S),

MoS$_2$-CdSe has four elements (i.e., Mo, S, Cd, and Se), and MoS$_2$-V-CdSe has five elements (i.e., Mo, S, Cd, Se, V, and O). The plasma formation with elemental composition follows the order as MoS$_2$-V-CdSe> MoS$_2$-CdSe> MoS$_2$ under similar HP intensities; thus, the results also reflect the same order in harmonics yields (Figure 2c).

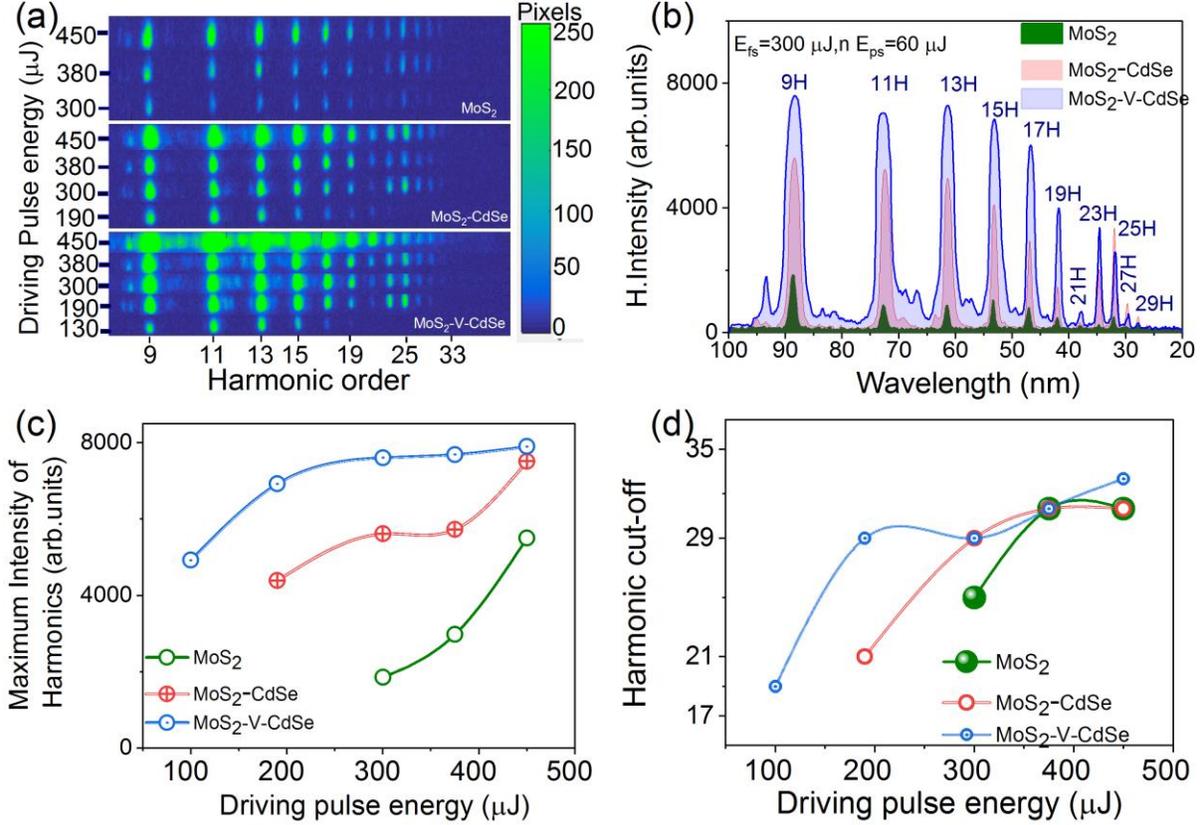

**Figure 3** (a) Raw HHG spectra for different fs DP energies (b) Comparison of HHG spectra for three samples at $E_{fs}$=300 µJ, $E_{ps}$=60 µJ. (c) maximum intensities of the harmonics at variable DP energies, and (d) harmonic cut-offs at different $E_{fs}$.

The harmonic cut-off was increased for MoS$_2$-V-CdSe due to an increase in their pondermotive potential ($U_p$), i.e., the average kinetic energy of accelerated electrons. The harmonic yield enhancement is related to the wavelength of driving pulses ($\lambda$) as $\sim \lambda^{-x}$, where $x \approx 5-6$.[14–16] Conversely, the harmonic cut-off is proportional to $\lambda^2$. According to the energy cut-off law, the maximum photon energy that can be emitted upon recombination is $E_{cut-off} = I_p + 3.17 U_p$, where $I_p$ is an ionization potential, $U_p = 9.33 \times 10^{-14} I\ (W.cm^{-2})\ \lambda^2\ (\mu m)$ is a ponderomotive energy of the electron, $I$ is a laser intensity, and $\lambda$ is a wavelength.[17] In this samples cases, this cut-off law agrees well with experimental findings at lower driving pulse energies. Notice that cutoff at lower driving pulse energies(intensities) is linearly depends on the energy (intensity) of laser pulse as shown in Figure 3d, the cutoff saturates with the growth of DP intensity. Probably, this is due to the increase in phase mismatch in plasma caused by the presence of free electrons, which diminish the yield of the higher order harmonics thus that they cannot be detected. The first ionization potentials of Mo, S, Cd, Se, V, and O are 7.09, 10.36, 8.99, 9.75, 6.74, and 13.61 eV, and the second ionization potentials are 16.16, 23.33, 16.90, 21.19, 14.66, and 35.11 eV, respectively. Among

the elements Mo, S, Cd, Se, V, and O, Vanadium (V) has the lowest ionization potential. Therefore, during the ablation of $MoS_2$-V-CdSe, could achieves the harmonics emission at a lower DP energy of 130 µJ.

*HHG from plasma using two-color pump (800 nm+400 nm):*

In the present case for the two-color pump (800 nm+ 400 nm) we have used the 0.2 mm thick BBO crystal for the conversion of second harmonic (2ω) from fundamental (ω) beam the obtained harmonic spectra for three samples were shown in Figure 4. Figures 4a and 4b depict the raw HHG spectra and corresponding line spectra for the TCP, respectively. All samples show a similar tendency of harmonic intensity and the increased cut-off for a two-color pump (TCP) at $E_{fs}$=450 µJ ($4.1\times10^{14}$ W/cm$^2$ ) $E_{fs,\ SH}$=36 µJ ($3.2\times10^{13}$ W/cm$^2$), $E_{ps}$=120 µJ ($7.12\times10^{9}$ W/cm$^2$) compared to single-color pump (SCP). In the case of TCP, the harmonic spectra include the both (even and odd) harmonics i.e. odd (9H, 11H, 13H, 15H, 17H, …, 29H) and even (8H, 10H, 12H, 14H, 16H,…, 30H) orders. The 10H has maximum intensity, similar to all even harmonics corresponding to 2(2n+1) orders (14H, 18H, 22H, 26H, 32H, …). Those harmonics correspond to the odd harmonics of 400 nm wave and predictable exceed the 4(n+1) odd harmonics (12H, 16H, 20H, 24H, 28H, …). Overall, the intensity of even harmonics is stronger than the intensity of odd harmonics for TCP due to the strong confinement of a short quantum path component with a higher ionization rate and a denser electron wave packet compared to the SCP [18,19].

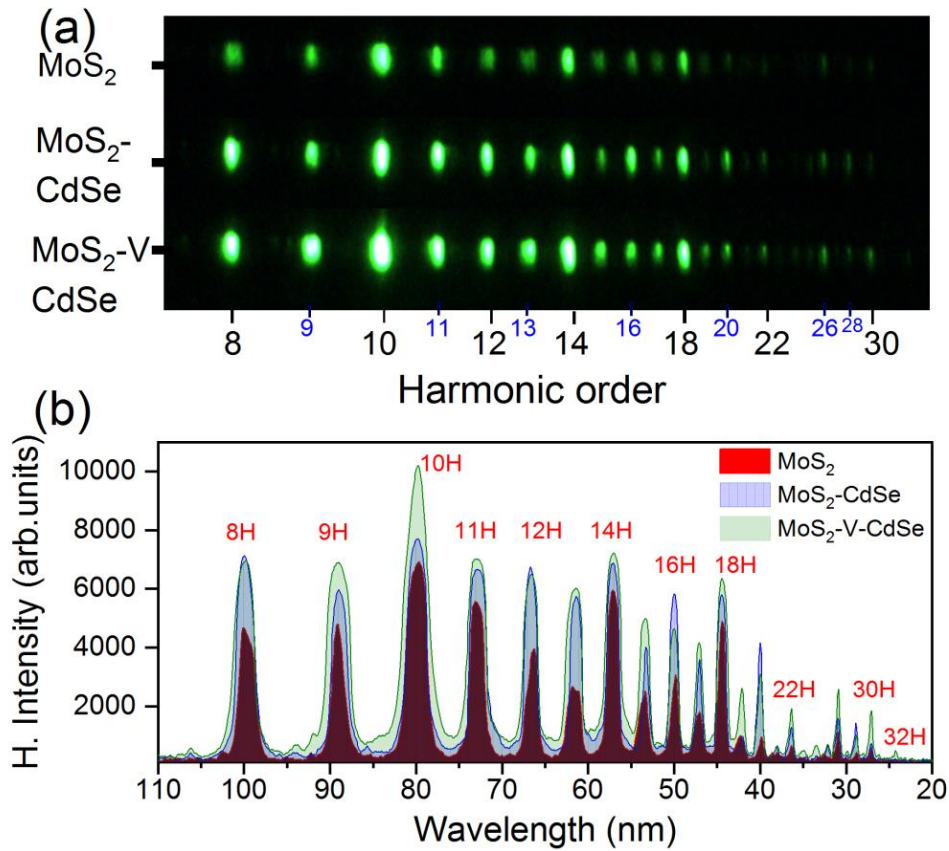

**Figure 4** (a) the raw HHG spectra and (b) corresponding line spectra for a two-color pump at $E_{fs}$=450 µJ ($4.1\times10^{14}$ W/cm$^2$), $E_{fs,\ SH}$=36 µJ ($3.2\times10^{13}$ W/cm$^2$), $E_{ps}$=120 µJ ($7.12\times10^{9}$ W/cm$^2$).

In the case of 0.2-mm thick BBO, the relative positions of the fundamental and SH pulses in the probing two-color pulse, as well as their respective pulse durations of 800 nm and 400 nm, were provided in [20]. The existence of the relatively weak second harmonic (SH) field leads to the presence of even harmonics, together with the odd harmonics $(2n + 1)\omega$ of the fundamental radiation generating at the frequency $\omega$ in the case of lower thickness BBO crystal. If the thickness of BBO crystal is increased which leads to increase in the walk off length (delay will increase) between SH and fundamental field; in this case the generation of the odd orders from each field, $(2n + 1)\omega$ and $(2n + 1)(2\omega)$ could be possible. Such discrepancy between the two pumps has been proved in [21] in the case of TCP using the 0.7 mm thick BBO crystal. In that case, the thick crystal significantly limited the overlap of two pulses in the producing media by introducing a substantial delay between them. In LIP, delay will be small enough to overlap two pulses if thin BBO crystals are used. Similarly, even though the SH conversion efficiency is much lower than in thick crystals, the harmonics of all orders of fundamental radiation may be seen in the HHG spectrum (Figure 4).

## Conclusions

We demonstrated the generation of high-order harmonics in the plasmas containing $MoS_2$-CdSe and $MoS_2$-V-CdSe and compared them with pristine $MoS_2$ nanosheets using 35 fs DP at SCP and TCP conditions of laser-plasma interaction. The LIPs were produced by HP at 1064 nm (5 ns) and 800 nm (200 ps). The harmonics intensities and cut-offs were increased for $MoS_2$-CdSe and $MoS_2$-V-CdSe LIPs compared to the case of the pure $MoS_2$ nanosheets-contained plasma (30H and 18H cut-ffs, correspondingly, in the case of two-color pump). Our studies have demonstrated that the additional presence of CdSe and $V_2O_5$ quantum dots leads to increase of the density of plasma components. As a result, these plasma components lead to enhance the emission of intense XUV radiation allowing further applications in attosecond physics and nonlinear spectroscopy.